\documentclass[11pt]{article}

\usepackage[numbers]{natbib}

\usepackage[vlined,linesnumbered,ruled,resetcount]{algorithm2e}
\usepackage[colorlinks,linkcolor=magenta,filecolor=blue,citecolor=blue,urlcolor=blue]{hyperref}%
\usepackage[top=1in, left=1in, right=1in, bottom=1in]{geometry}
\usepackage{pifont}

\usepackage{authblk}
\usepackage{enumitem}

\usepackage{times}
\usepackage{latexsym}

\usepackage{graphicx}   
\usepackage{caption}    
\usepackage{subcaption} 

\usepackage[T1]{fontenc}

\usepackage[utf8]{inputenc}

\usepackage{microtype}

\usepackage{inconsolata}

\usepackage{graphicx}
\usepackage{enumitem}

\newcommand{\BB}{\vspace*{-\medskipamount}}
\newcommand{\BBB}{\vspace*{-\bigskipamount}}

\usepackage{array}
\usepackage{booktabs}
\usepackage{multirow}
\usepackage{amssymb}
\usepackage{pifont}

\newcommand{\goodname}{\textsf{ScaleLLM}}
\usepackage{threeparttable,booktabs}
\usepackage{amsmath}

\makeatletter
\newcommand*{\RN}[1]{\expandafter\@slowromancap\romannumeral #1@}
\makeatother

\makeatletter
\newcommand{\printfnsymbol}[1]{%
  \textsuperscript{\@fnsymbol{#1}}%
}
\makeatother

\title{ScaleLLM: A  Resource-Frugal LLM Serving Framework by Optimizing End-to-End Efficiency}

\author[1]{Yuhang Yao}
\author[1]{Han Jin}
\author[1]{Alay Dilipbhai Shah}
\author[1]{Shanshan Han}
\author[1]{Zijian Hu}
\author[1]{Yide Ran}
\author[1]{Dimitris Stripelis}
\author[1]{Zhaozhuo Xu}
\author[1]{Salman Avestimehr}
\author[1]{Chaoyang He}

\affil[1]{TensorOpera Inc.}
\affil[ ]{{\texttt{\{yuhang,raphael,alay,shanshan,zjh,ryan,dimitris\\zhaozhuo,ch,avestimehr\}@tensoropera.com}}}

\begin{document}
\date{}
\maketitle
\begin{abstract}
    Large language models (LLMs) have surged in popularity and are extensively used in commercial applications, where the efficiency of model serving is crucial for the user experience. Most current research focuses on optimizing individual sub-procedures, \textit{e}.\textit{g}. local inference and communication, however, there is no comprehensive framework that provides a holistic system view for optimizing LLM serving in an end-to-end manner. In this work, we conduct a detailed analysis to identify major bottlenecks that impact end-to-end latency in LLM serving systems. Our analysis reveals that a comprehensive LLM serving endpoint must address a series of efficiency bottlenecks that extend beyond LLM inference. We then propose~\goodname, an optimized system for resource-efficient LLM serving.
Our extensive experiments reveal that with 64 concurrent requests, ~\goodname~ achieves a 4.3$\times$ speed up over vLLM and outperforms state-of-the-arts with 1.5$\times$ higher throughput\footnote{\url{https://tensoropera.ai/prod/model/mistralai/ScaleLLM-Mixtral-8x7B}}.


\end{abstract}

\section{Introduction}
Large language models (LLMs) have significantly changed the field of natural language processing and have been widely used in commercial applications. 
However, serving LLMs effectively remains challenging due to system latency, query concurrency, and computational resources constraints. 
LLM applications are typically deployed as online services where users expect real-time responses, while any delay can impact user experience, making low latency to be crucial. Also, the computationally intensive nature of LLMs, which involve inference with billions of parameters, requires substantial computational resources. 
Moreover, achieving scalability to handle multiple concurrent requests without performance degradation further complicates the serving process.

Latency in LLM serving primarily arises from the processing at the serving engine as well as the gateway. 
The serving engine is the core component responsible for executing the LLM inference tasks. It optimizes resource allocation to handle the intensive computational workload of LLMs to efficiently utilize computational resources, such as GPUs. 
The gateway manages communication between clients (\textit{e}.\textit{g}., end-users or applications) and LLM instances. It handles incoming requests, directs them to the LLM instances, and ensures that responses are returned correctly and efficiently. 

\begin{figure*}[ht]
    \centering
    \includegraphics[width=0.95\linewidth]{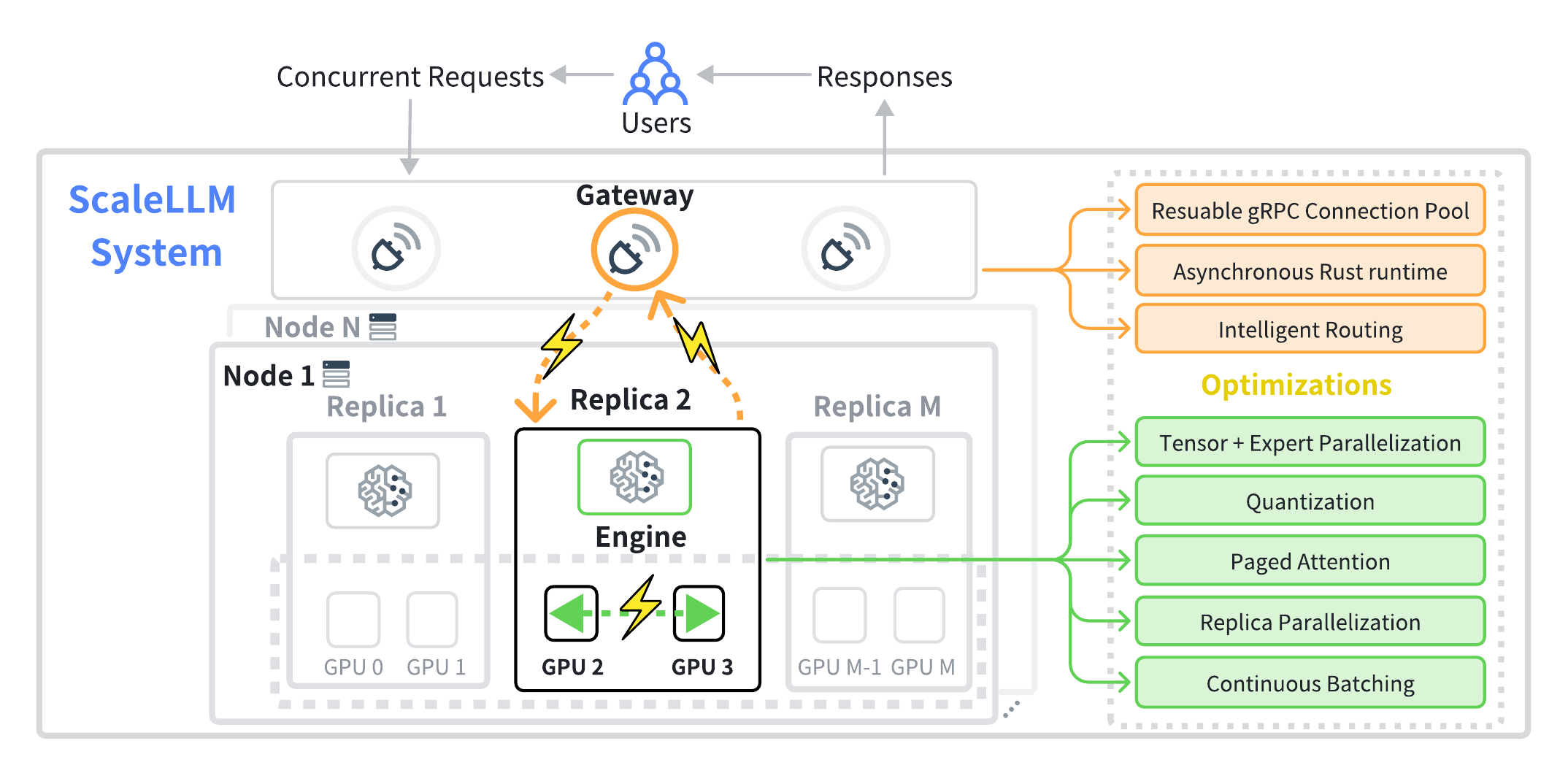}
    \caption{Overview of~\goodname~Serving System.~\goodname~ provides an optimized gateway for balancing workloads of user requests to different inference replicas and an efficient serving engine for promptly response with high concurrent requests.} 
    \label{fig:platform_local_tiem_diagram}
\end{figure*}

Existing research focuses on optimizing individual subprocedures of LLM serving, especially accelerating local inference speeds~\cite{dao2022flashattention, VLLMAI, TensorRTLLM}. However, in commercial LLM applications, end-to-end latency, introduced from functionalities of the gateway, becomes the most significant bottleneck. 
Meanwhile, commercial LLM applications have specific requirements on serving, which directly accessing a single LLM instance fails to address. 
In practice, commercial LLM applications must satisfy several critical requirements for efficient and reliable inference: \textit{i}) \textit{fault tolerance}: there must be replicas of LLMs to ensure that the serving system can select appropriate replica upon receiving requests under a specific resource constraint, thereby maintaining service reliability even when individual replica instance fails; 
\textit{ii}) \textit{inference control}: the serving system should manage the inference process to ensure that the models are accessed with authentication and can produce responses that are appropriate and safe while adapting to different user demands; 
\textit{iii}) \textit{low latency}: to ensure the user experience, the serving system should process inferences efficiently and deliver responses in real time; 
\textit{iv}) \textit{concurrency}: small batch sizes and high throughput for individual requests become impractical in real-world LLM services such as ChatGPT, where the queries can be frequent, \textit{e}.\textit{g}., with queries per second (QPS) often exceeding 200~\cite{chatgpt_statistics}; 
\textit{v}) \textit{frugal computational resource usage}: given the substantial computational demands, optimizing resource utilization is crucial to prevent excessive costs and ensure the reliable operation of the serving system. 
Thus, a comprehensive LLM serving system must balance computational efficiency, concurrency, and latency to manage the high volume of requests. 

To address the efficiency of LLM serving comprehensively, we present~\goodname, an optimized LLM serving system, as well as an end-to-end measurement, to meet real-world requirements of commercial LLM applications. As shown in Figure~\ref{fig:platform_local_tiem_diagram}, to address different challenges in commercial LLM applications, ~\goodname~optimizes two crucial modules, including \textit{i}) a Routing Module that efficiently does replica level load balancing and data transmission;
and \textit{ii}) a strong LLM engine to inference promptly with high concurrent requests. 
Additionally,~\goodname~features a Safety Module for authentication, rate limiting, and sensitive content detection as well as an Observability Module that persists metrics to local disk to adhere to standard requirements of real-world production systems. 
Our contributions are summarized as follows.

\begin{itemize}[nosep,leftmargin=*]
    \item We go beyond optimizing the latency of LLM inference and measure the end-to-end time and resource cost of maintaining an LLM serving endpoint. Moreover, we present
    a breakdown of the end-to-end LLM serving endpoint to showcase the overhead introduced in each component.
    
    \item We optimize LLM serving for both the local inference and the gateway, and provide a recipe for efficient LLM serving frameworks for commercial applications. 
    Specifically, instead of random selection, we evaluate different gateways in \S\ref{sec:optimize_router}, and choose Rust as the backend due to its superior performance in terms of latency, concurrency handling, and resource efficiency.
    
    \item Extensive experiments highlights that with 64 concurrent requests, \goodname~achieves a 4.3$\times$ speed up over vLLM and outperforms the state-of-the-arts with 1.5$\times$ higher throughput~\cite{FireworksAI,TogetherAI}. 
    \item Lastly, we synthesize our insights and findings from extensive experiments into the \textit{blueprint design} of a dynamic inference load balancing system engineered to adapt to varying workloads to address the critical requirements of the contemporary production environments.
\end{itemize}

\section{Related Work}
Many pre-trained open LLMs have been released since last year, where the most commonly used models including Mixtral 8x7B~\cite{jiang2024mixtral} and Llama-3~\cite{touvron2023llama}). Such open-source models motivate the industry to build public LLM-serving endpoints~\cite{TogetherAI, FireworksAI} and empower researchers to work on speeding up the inference speed. FlashAttention~\cite{dao2022flashattention} is proposed to approximate the attention calculation to reduce memory usage with fast computation. By representing the weights and activations with low-precision data types, Model Quantization~\cite{lin2024awq,liu2024kivi} is also widely adopted to reduce memory and computation costs. 

During LLM serving, the key-value cache (KV cache) memory for each request is huge and grows and shrinks dynamically, Page attention~\cite{kwon2023efficient} is proposed for efficient management KV cache memory blocks with exact model computation. Built on top of PagedAttention, vLLM~\cite{VLLMAI} is proposed as a high-throughput distributed LLM serving engine that aims to increase GPU utilization and hence speeds up the throughput of LLM serving. TensorRT-LLM~\cite{TensorRTLLM} provides industrial-level integration of these state-of-the-art optimization methods with Python and C++ runtimes to perform inference efficiently on NVIDIA GPUs.

However, these serving engines primarily focus on accelerating local LLM computation, neglecting other crucial components such as gateway and routing. To the best of our knowledge, our proposed ScaleLLM is the first to offer an end-to-end latency measurement and optimization specifically for resource-efficient LLM serving.

\section{Benchmark LLM Serving Solutions}
We first provide the end-to-end system breakdown of serving latency in \S\ref{sec:breakdown}, then provide the benchmark results of baselines in \S\ref{sec:benchmark_baselines}.


\subsection{System Breakdown}\label{sec:breakdown}
To optimize the user's experience with low latency, there are two components to be focused on.



\noindent\textbf{Replica Router.} In practical applications, the serving endpoint is not a single instance but consists of multiple replicas and schedulers to facilitate load balancing. The router functions as a crucial module that mediates request and response transformation between the engine and the end user. Given the high concurrency of user requests, the router typically operates under significant pressure.

\noindent\textbf{Inference Engine within Replica.} A replica represents the smallest unit of resource allocation and is designed to be homogeneous. Each replica houses an instance of the inference engine, utilizing one or more GPUs with a specific parallelism pattern, such as tensor parallelism or process parallelism. 

\begin{figure}[ht]
    \centering
    \begin{subfigure}{0.47\textwidth}
        \centering
        \includegraphics[width=\textwidth]{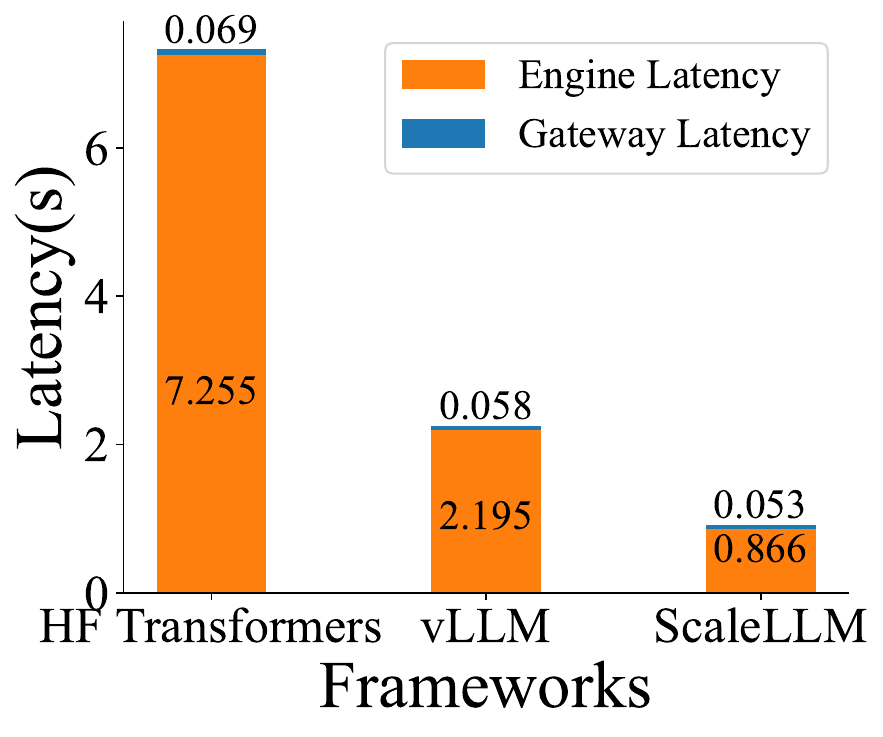}
        \caption{Concurrency: 4.}
        \label{fig:latency_vs_concurrency_4}
    \end{subfigure}%
    \begin{subfigure}{0.47\textwidth}
        \centering
        \includegraphics[width=\textwidth]{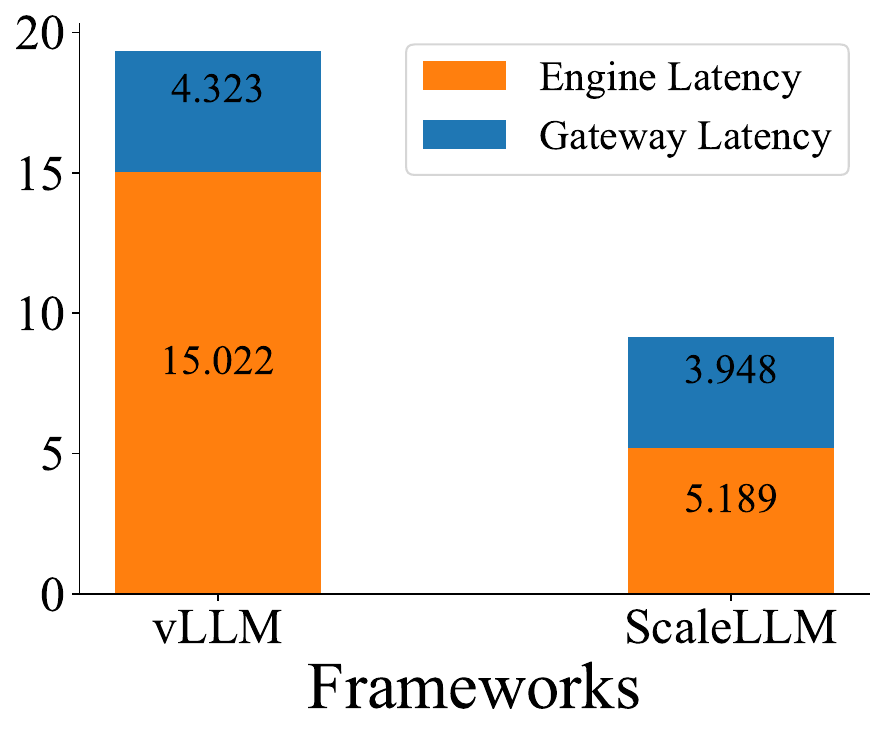}
        \caption{Concurrency: 256. }
        \label{fig:latency_vs_concurrency_256}
    \end{subfigure}%

    \caption {Comparisons with the two baseline solutions. ScaleLLM is applied without gateway optimization.}
    \label{fig:benchmark}
\end{figure}

\subsection{Performance of Baseline Solutions}\label{sec:benchmark_baselines}
For the routing gateway, FastAPI is widely adopted due to its user-friendliness and ease of setup. For the serving engine, there are two baselines, including Huggingface Transformer~\cite{wolf2019huggingface} and vLLM~\cite{kwon2023efficient}. Benchmark results in Figure~\ref{fig:benchmark} indicate that with 4 concurrent requests, engine latency is the primary bottleneck. However, at 256 concurrent requests, the gateway latency becomes the predominant bottleneck.

\section{Optimizations}\label{sec:optimization}
This section discusses the optimization goal, then decompose the latency into engine latency and gateway latency, and optimize each component. 

\noindent\textbf{Optimization Goal. } 
Our goal is leveraging various optimization techniques on both the inference engine and the replica router to improve the end-to-end serving performance. 
The inference engine are applied with different frameworks and optimization methods to increase the throughput and decrease the latency.
For the replica router, we break down the latency to engine latency and gateway routing latency. The goal is to decrease the engine latency, especially when the concurrency is high.

\subsection{Optimize Inference Engine}\label{sec:optimize_engine}
We mainly focus on optimizing the Mixture of Experts ~\cite{jiang2024mixtral}  LLMs that are being widely used nowadays.

\noindent\textbf{Model Parallelization.}
We utilize parallel processing across multiple GPUs to accommodate models with multiple experts (MoEs), as the model may not fit within the memory of a single GPU. As shown in Figure~\ref{fig:tp_ep_parallelism}, TensorRT engine \cite{TensorRTLLM} offers three approaches for achieving parallelism, including Tensor Parallel, Expert Parallel, and a hybrid of the two.  Tensor parallelism (TP) is a method for distributing a model's computation across multiple GPUs by splitting tensors into non-overlapping pieces, which allows different parts of the tensor to be processed simultaneously on separate GPUs. Expert Parallelism (EP), on the other hand, distributes experts of an MoE across GPUs. We found that a hybrid mode for balancing TP and EP can be 1.5$\times$ faster over the original TP solution; see \textbf{Exp4} in \S\ref{sec:exp} for details.



\noindent \textbf{Model Quantization.}
During model inference, each parameter of original LLM model is stored as a float number with 32-bit (fp32), resulting in significant GPU memory consumption and slower inference speeds. However, applying quantization techniques using 16-bit (fp16) and 8-bit (fp8) floating point numbers can substantially reduce memory usage and accelerate inference speeds, while maintaining nearly the same model accuracy as fp32~\cite{liu2024kivi,lin2024awq}.

\noindent \textbf{Continuous Batching and Batch Scheduler.}
To efficiently handle asynchronous user requests, we use a continuous batching strategy that batches requests for simultaneous processing by the engine. This method addresses variability in user input characteristics, such as input length, which can cause inefficiencies in static batching.
Furthermore, our experiments with scheduling policies revealed that setting policy to max utilization, when in-flight sequence batching is enabled, significantly enhances GPU utilization by processing maximum number of requests per iteration. However, this aggressive approach may require pausing requests if KV cache size limit is reached, a trade-off to consider in production systems.



\noindent \textbf{Other Optimizations.}
We adopt Flash Attention~\cite{dao2022flashattention} for operator fusion and Paged Attention \cite{kwon2023efficient} to boost the performance

\begin{figure}[ht]
    \centering
    \includegraphics[width=0.98\linewidth]{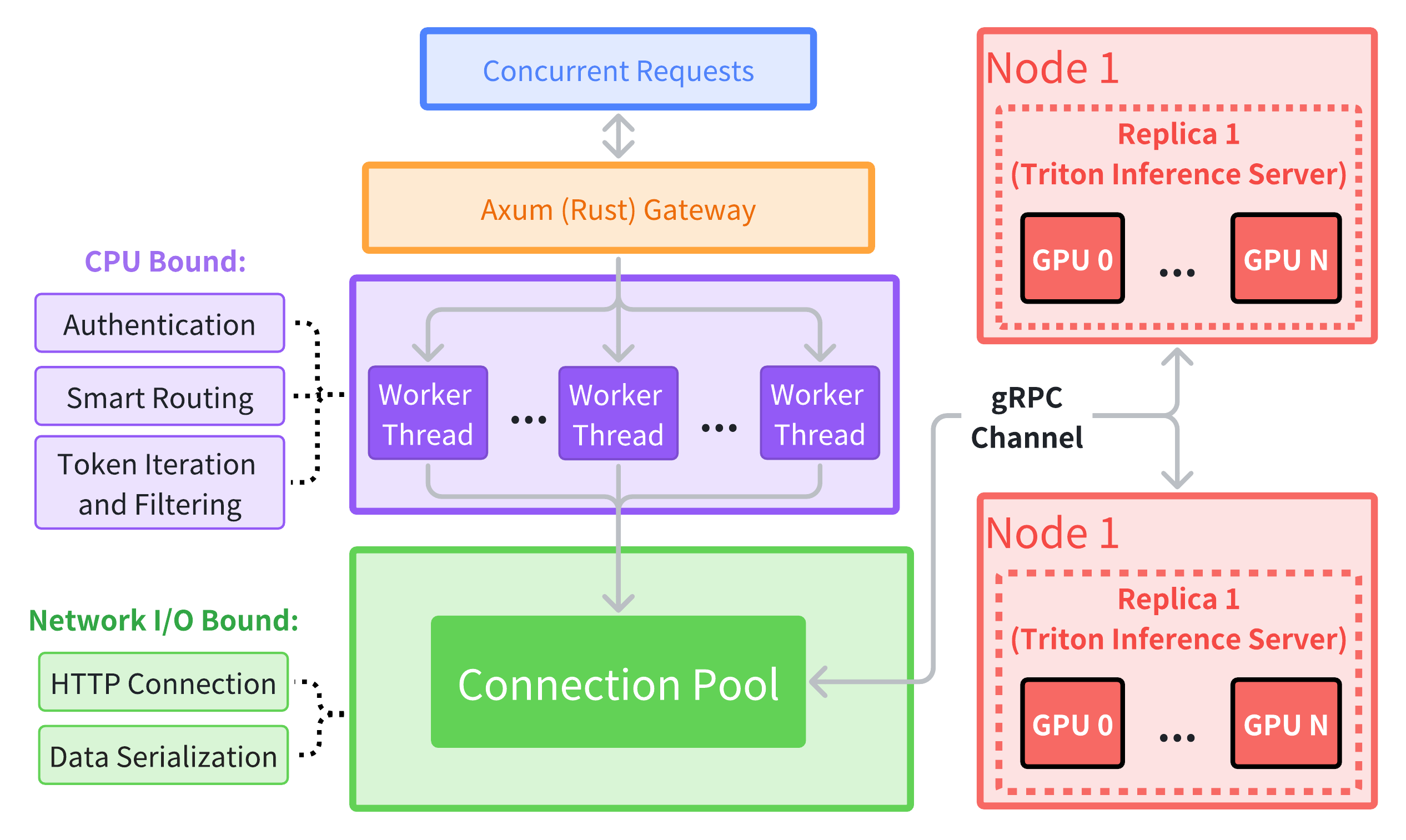}
    \caption{ScaleLLM Gateway Architecture}
    \label{fig:rust_gateway}
\end{figure}

\subsection{Optimize Replica Router}
\label{sec:optimize_router}
To effectively manage high concurrent requests, the gateway must exhibit superior performance in handling extensive Network I/O, database I/O, and CPU-intensive operations, including authentication processes, routing algorithms, and token filtering for security purposes. The efficient execution of these resource-bound tasks is critical, as they significantly impact the system's overall latency and throughput. Optimizing the gateway's capacity to handle these diverse and demanding operations is essential for maintaining system performance and scalability under high-load conditions.
To address these requirements, we replace the baseline router framework, which is based on FastAPI (Python), with Axum (Rust). In term of transaction protocol, we migrate from HTTP/1.1 to the gRPC protocol. The architecture is shown in Figure~\ref{fig:rust_gateway}.

\noindent \textbf{CPU Bound Job Optimization.}
For CPU bound jobs,  the FastAPI gateway in the baseline implementation is constrained by the Global Interpreter Lock (GIL), which limits its ability to utilize multiple CPU cores effectively. We refactor the gateway using Tokio \cite{Asynchronous_Runtime} for multi-task execution across multiple worker threads and Axum \cite{Web_Framework} for web development. 

\noindent \textbf{Network I/O Bound Job Optimization.}
We implement a gRPC connection pool based on Tonic \cite{gRPC_Framework}, a robust and efficient gRPC framework. This approach allows new requests to reuse existing connection channels, thereby reducing connection establishment overhead. Additionally, by utilizing Protocol Buffers for data serialization, we further decreased associated costs.


\section{Experiments}\label{sec:exp}

\begin{figure}[ht]
\BB
    \centering
    \includegraphics[width=0.7\linewidth]{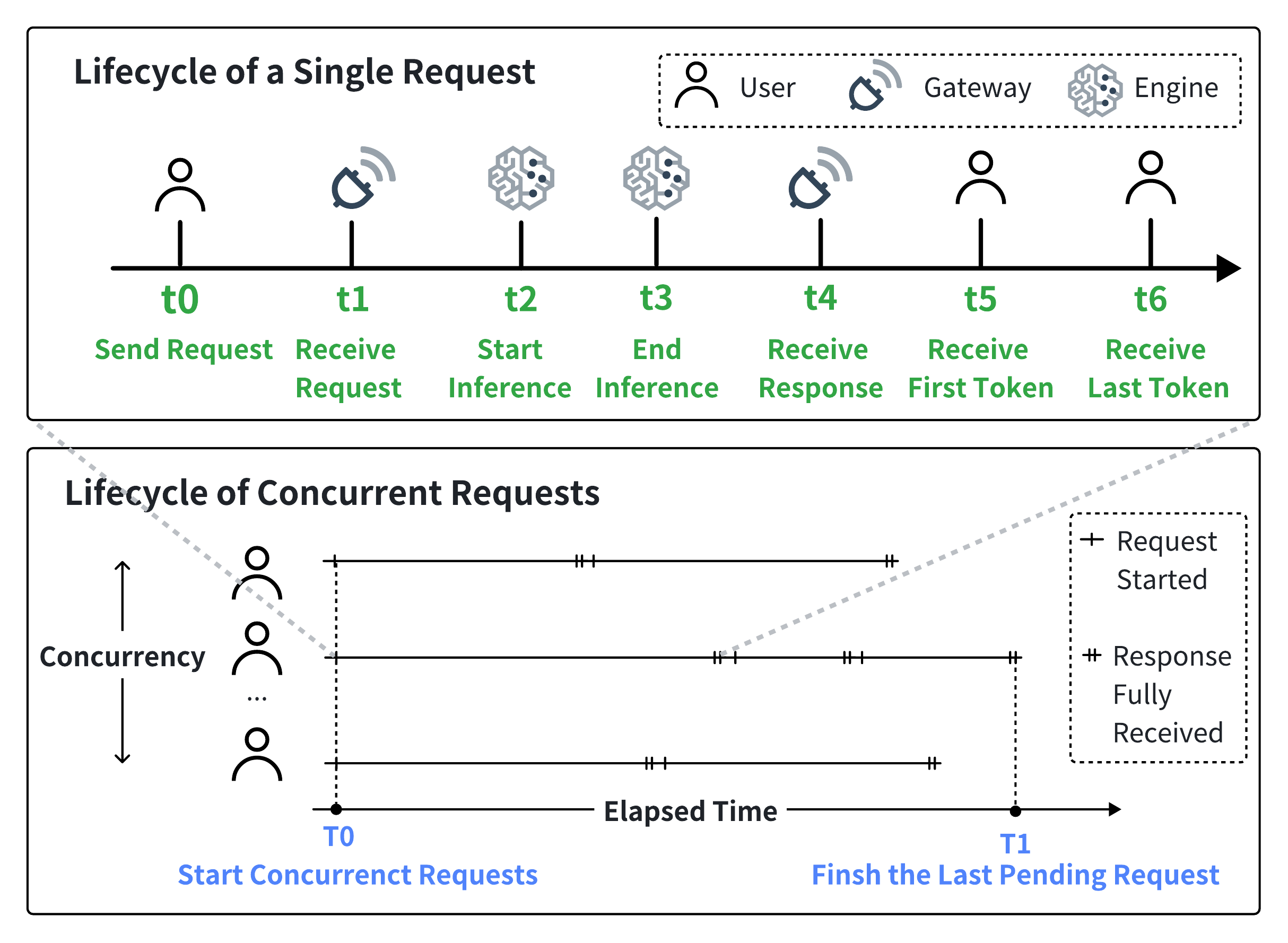}
    \caption{Lifecycle of Concurrent and Single Request}
    \label{fig:Experiment Glossary}
\end{figure}

\textbf{Experimental settings. }
We employ 8 NVIDIA DGX H100 GPUs, connected via 18 NVLink links, each providing a bandwidth of 26.562 GB/s.
We select Mistral 8x7B~\cite{jiang2024mixtral} as the inference LLM, and set the maximum tokens generation length to 512, the temperature to 0.5, and the  top-p parameter to 0.7. We optimize the ScaleLLM engine based on TensorRT-LLM~\cite{TensorRTLLM}. Our evaluations use OpenOrca dataset~\cite{OpenOrca} that contains question-response pairs for LLMs, as well as predefined system prompts. We simulated user's behavior of submitting prompt in OpenAI API format~\cite{openai_api} to the system, in a concurrent and continuous manner. Figure~\ref{fig:Experiment Glossary} illustrates the typical lifecycle of concurrent requests in comparison to one request .

\noindent\textbf{Compared Endpoints. } 
We utilized several endpoints for comparisons, including
\textit{i}) \textit{\textbf{Huggingface}} \textit{\textbf{Endpoint}} that is deployed with Huggingface transformer~\cite{wolf2019huggingface} and FastAPI gateway; \textit{ii}) \textit{\textbf{vLLM Endpoint}} that is deployed with vLLM~\cite{VLLMAI} and FastAPI gateway; and \textit{iii}) \textit{\textbf{Fireworks and Together AI Endpoints}}~\cite{FireworksAI,TogetherAI}.

\subsection {Evaluation Metrics}\label{sec:eval_metrics}
We define metrics to evaluate the efficiency of LLM serving frameworks. To explain the definitions clearly, we illustrate different stages of LLM inference in Figure~\ref{fig:Experiment Glossary}. 
For the rest of \S\ref{sec:eval_metrics}, we denote $t_0$ as the timestamp the user submits a request, $t_1$ as the timestamp for the router to receive that request, $t_2$ as the start time for engine's local inference, $t_3$ as the engine finished the inference, $t_4$ as the time gateway received the response from engine, $t_5$ as the time for the user to receive the first token, and $t_6$ as the timestamp that they receive the full output.

\noindent\textbf{\# of Concurrent Requests}: The upper bound of the number of ongoing requests at a single moment.

\noindent\textbf{\# of Requests:} In order to fulfill system during a elapsed time period, this number set to be $20 \times c$  where c is the number of concurrent requests.

\noindent\textbf{Average Latency}: The average waiting time for a user to see the full output, computed as $t_5-t_0$.\\
\noindent\textbf{Gateway Latency}: The time cost for processing and routing requests and LLM responses between the user and the inference engine, defined as ${(t_2-t_0) + (t_5-t_3)}$, where $t_2-t_0$ is the time for processing and routing a user request to the inference engine, and $t_5-t_3$ is  the time for transferring the response from the engine to the user.
\\
\noindent\textbf{Engine Latency:} The time for the engine to process a local inference, computed as $t_3-t_2$.
\\
\noindent\textbf{Throughput:} 
The number of tokens that the whole system generates within a certain time frame, computed as $\frac{N_t}{T_1 - T_0}$, where $N_t$ is the number of generated tokens, $T_1$ is the timestamp to finish the last request, and $T_0$ is the time that the concurrent requests start.

\noindent\textbf{Time to First Token (TTFT):} The elapsed time between user to submit a new request and to receive the first token, computed as ${t_4-t_0}$. \\
\noindent\textbf{Time Between Tokens (TBT):} The average wait time to the next generated token after the first generated token, computed as 
$\frac{(N_g - 1)}{(t_6-t_5)}$, where $N_g$ is the number of generated tokens for one request.

\subsection{Serving Performance Evaluation}
We first provide the comparison with the state-of-the-art endpoints, then make detailed comparison for non-streaming and streaming generation.

\begin{figure}[ht]
    \centering
    \includegraphics[width=0.86\linewidth]{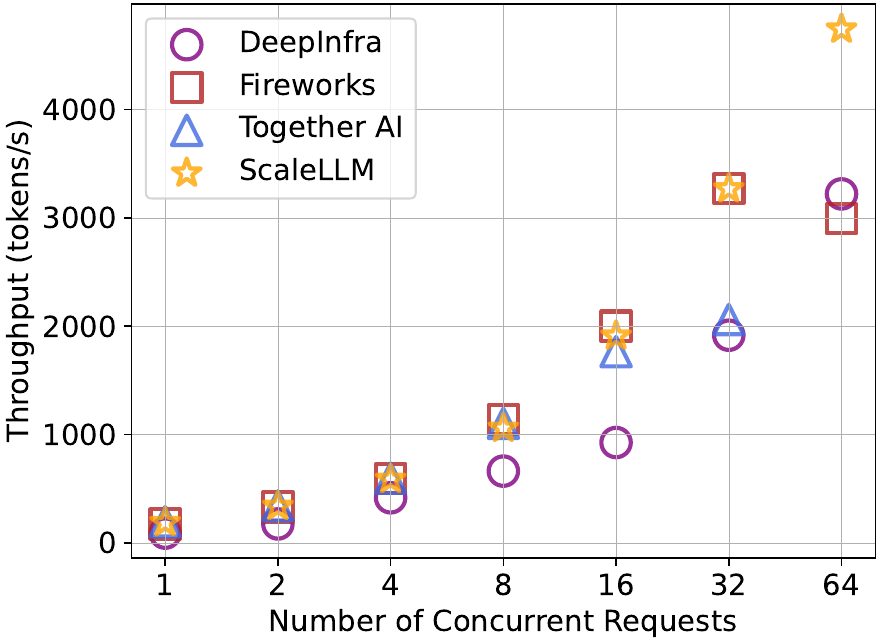}
    \caption{Endpoint Throughput Comparison.}
    \label{fig:Endpoint Throughput Comparison}
\end{figure}

\noindent\textbf{Exp1. Endpoints Throughput Comparison.} We compare the throughput of~\goodname~ against DeepInfra, Fireworks, and Together AI across different levels of concurrency. As shown in Figure~\ref{fig:Endpoint Throughput Comparison},~\goodname~performs comparably to other endpoints at lower concurrency levels. However, ~\goodname~ significantly outperforms the endpoints as the concurrency scales up, and surpasses all other endpoints by a huge margin for batch size 64.

\begin{figure*}[ht]
    \centering
    \includegraphics[width=1\linewidth]{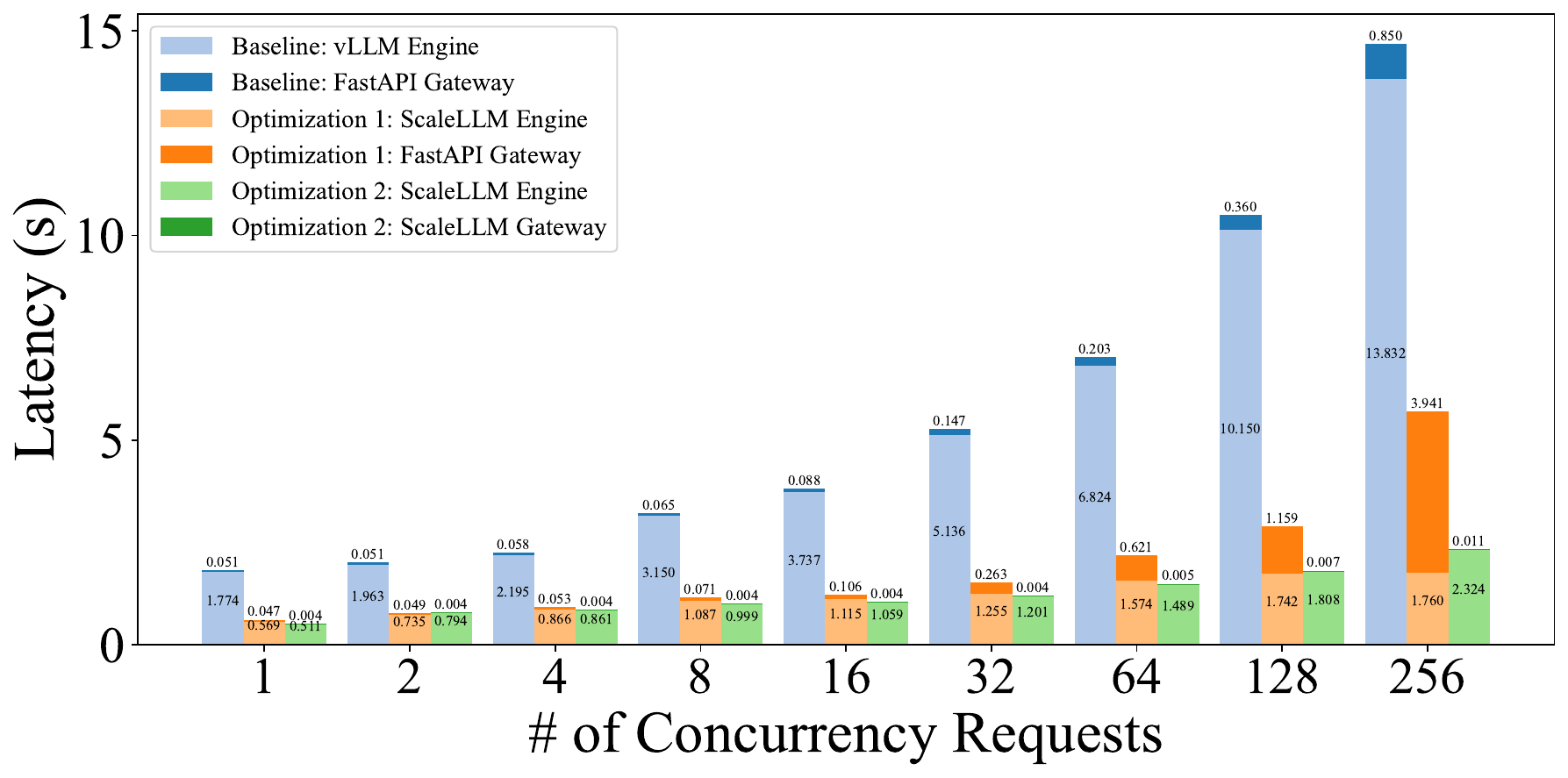}
    \caption{System latency vs \# of concurrent requests. }
    \label{fig:latency_vs_concurrency} 
\end{figure*}

\begin{figure*}[ht]
    \centering
    \includegraphics[width=1\linewidth]{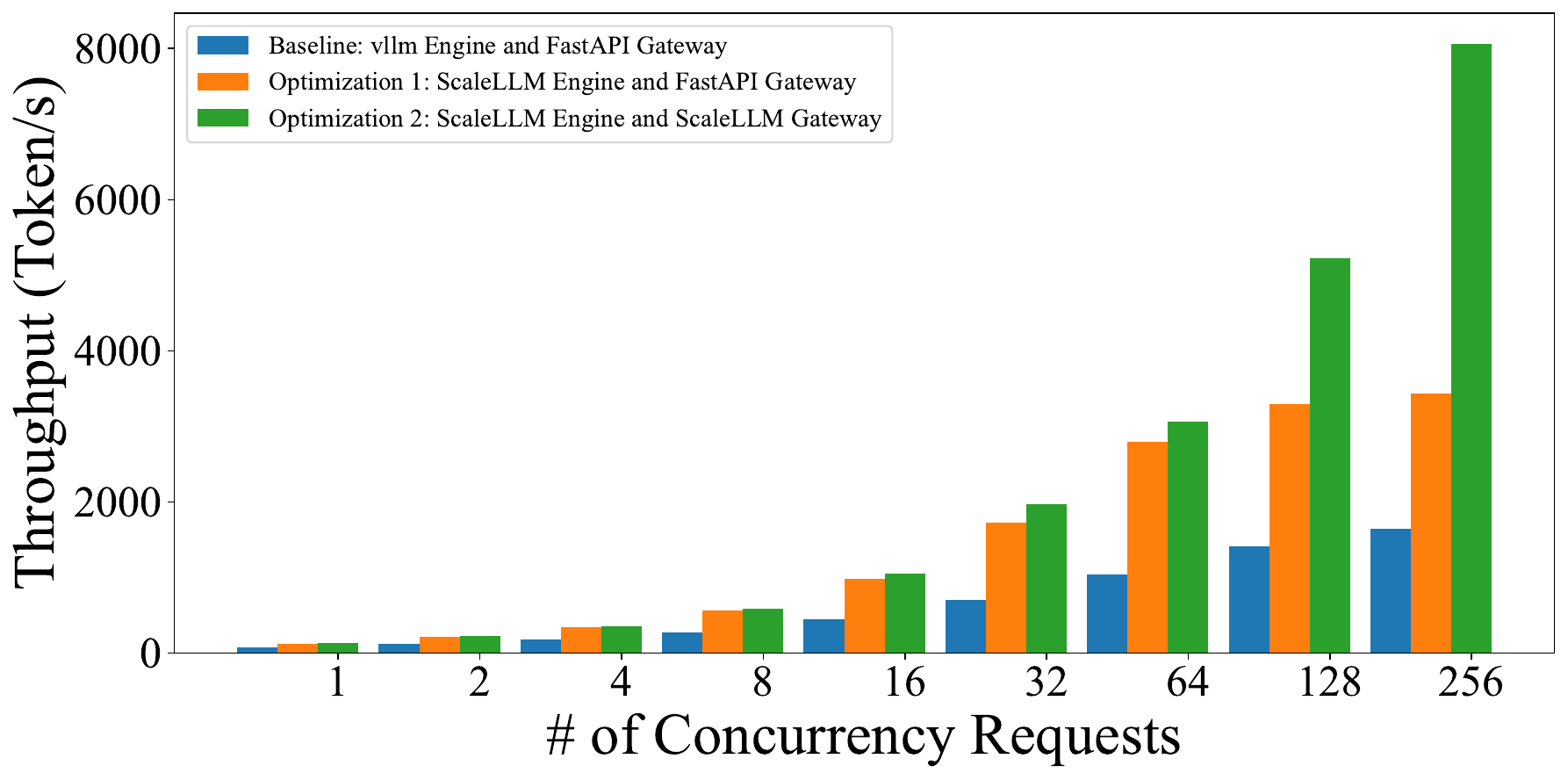}
    \caption{Throughput vs \# of concurrent requests. }
    \label{fig:system_throughput} 
\end{figure*}

\noindent\textbf{Exp2. Non-Streaming Generation Evaluation. }
\label{sec: Non-Streaming-Gen} We conducted a comprehensive latency breakdown evaluation for Mistral 8x7B running on two H100 GPUs, examining various levels of concurrent requests. The averaged latency decomposition is shown in Figure~\ref{fig:latency_vs_concurrency}. The result shows that with ~\goodname~, the engine latency is reduced compared to the baseline engine. However, at concurrency levels of 64/128/256, the baseline gateway latency increases when connected to the ~\goodname~ Engine, compared to its connection with Baseline engine, making it the new bottleneck. This is attributed to the baseline gateway's inability to keep pace with the ~\goodname~ Engine's generation speed due to CPU bound task and Network I/O task as mentioned in~\S\ref{sec:optimize_router}. However, we observe a significant reduction in latency upon swapping the baseline gateways out with ~\goodname~Gateway, indicating that ~\goodname~Gateway matches the engine's generation speed, thereby shifting the bottleneck back to engine.

We further evaluated the throughput differences between the ScaleLLM Engine and the vLLM Engine, as well as their integration with the FastAPI Gateway and the optimized ScaleLLM Gateway. The complete results (with Concurrency from 1 to 256) are illustrated in Figure~\ref{fig:system_throughput}. The findings indicate that engine optimization leads to significant improvements in throughput; Additionally, the optimization of the Gateway contributes to further notable performance enhancements, demonstrating the cumulative impact of both engine and gateway optimizations on overall system performance.





\begin{table*}[ht]
\centering
\caption{TTFT and TBT for end to end streaming requests. Smaller TTFT means faster response for the first token and smaller TBT means faster generation of tokens. Timeout: 90\% of the users' requests cannot complete in 60s.}
\label{tbl:ttft_tbt_comparison}
\small
\begin{tabular}{l|cc|cc|cc}
\toprule
\textbf{Concurrent} & \multicolumn{2}{c|}{\textbf{Huggingface Endpoint}} & \multicolumn{2}{c|}{\textbf{vLLM Endpoint}} & \multicolumn{2}{c}{\textbf{ScaleLLM}} \\ \cline{2-7} 
\textbf{Requests}            & TTFT/ms                 & TBT/ms                  & TTFT/ms                   & TBT/ms                  & TTFT/ms           & TBT/ms            \\ \hline
1                   & 315.6                   & 83.4                    & 48.4                      & 16.5                    & 25.0 (1.9x)       & 8.5 (1.9x)        \\
2                   & 637.2                   & 218.3                   & 51.9                      & 16.7                    & 25.3 (2.1x)       & 8.7 (1.9x)        \\
4                   & 1157.8                  & 506.4                   & 55.1                      & 21.1                    & 25.5 (2.2x)       & 10.4 (2.0x)       \\
8                   & Timeout                 & Timeout                 & 70.2                      & 30.1                    & 25.9 (2.7x)       & 12.2 (2.5x)       \\
16                  & Timeout                 & Timeout                 & 93.1                      & 38.3                    & 26.7 (3.5x)       & 13.4 (2.9x)       \\
32                  & Timeout                 & Timeout                 & 135.8                     & 50.1                    & 29.8 (4.5x)       & 14.6 (3.4x)       \\
64                  & Timeout                 & Timeout                 & 285.4                     & 70.8                    & 99.4 (2.9x)       & 16.5 (4.3x)       \\ 
\bottomrule
\end{tabular}
\end{table*}

\noindent\textbf{Exp3. Streaming Generation Evaluation.} To provide an intuitive perspective from the user's point of view, we compared the time to the first token (TTFT) and time between tokens (TBT) on~\goodname~with Huggingface Transformer and vLLM. In order to simulate the realistic user's waiting threshold, we set the timeout of generating all the tokens to be 60 seconds. The results in Table~\ref{tbl:ttft_tbt_comparison} shows that the HuggingFace Endpoint has the highest TTFT and TBT, where over 90\% of the user's requests get timeout after 60 seconds when the concurrency is 8. In contrary, vLLM has lower TTFT and TBT but ~\goodname~improved over 1.9$\times$ lower TTFT and TBT comparied with the vLLM Endpoint. 



\begin{figure*}[ht]
    \centering
    \begin{subfigure}{0.2485\textwidth}
        \centering
        \includegraphics[width=\textwidth]{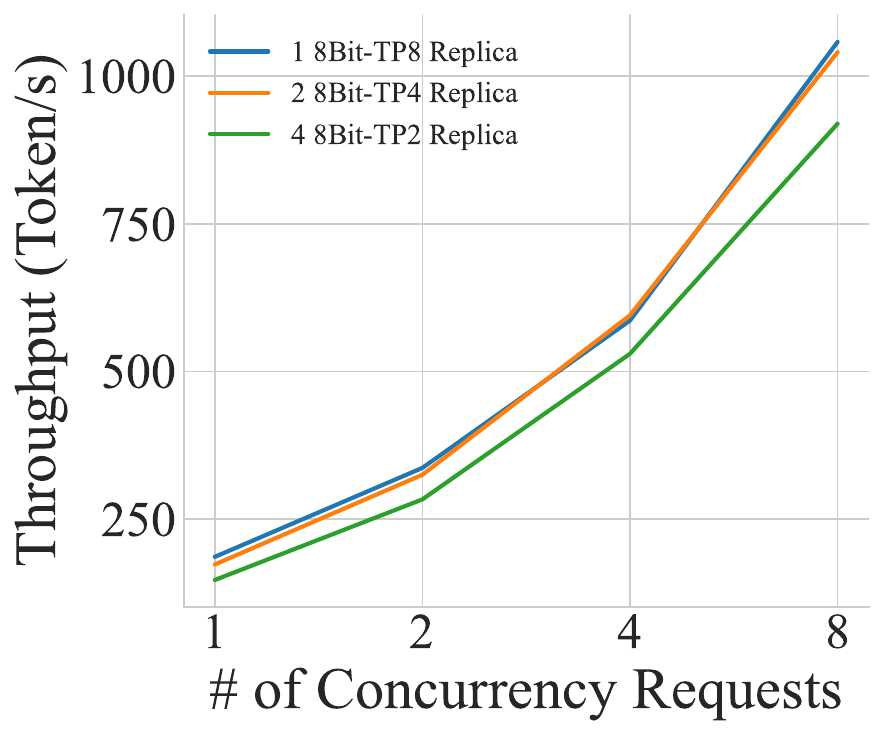}
        \caption{Low conc. with 1-4 replica.}
        \label{fig:low_concurrency_replica124}
    \end{subfigure}%
    \hfill
    \begin{subfigure}{0.2485\textwidth}
        \centering
        \includegraphics[width=\textwidth]{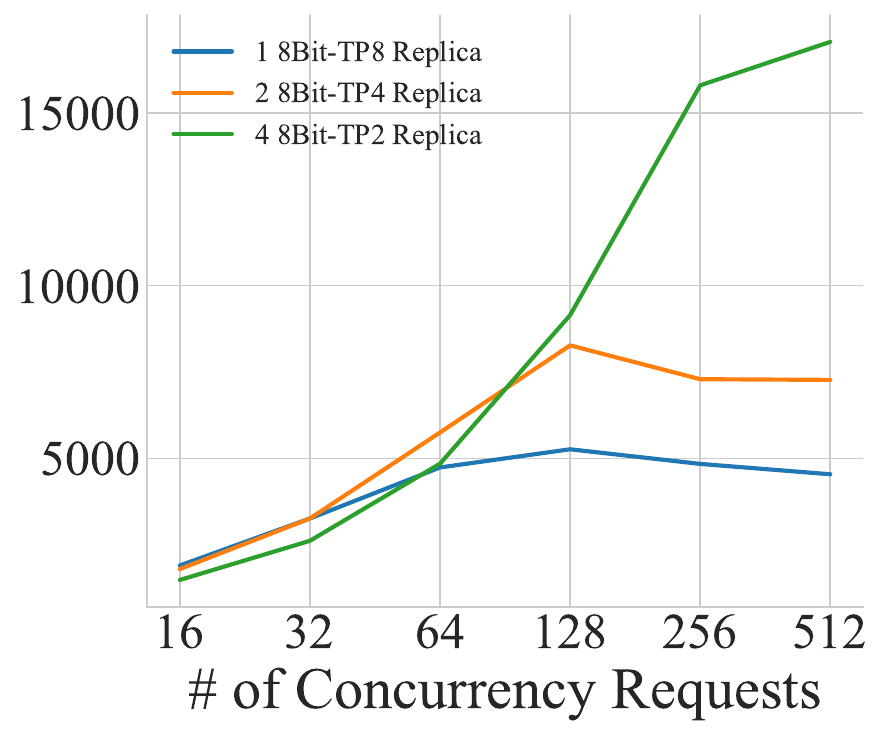}
        \caption{High conc. with 1-4 replica.}
        \label{fig:high_concurrency_replica124}
    \end{subfigure}
    \hfill
    \begin{subfigure}{0.2485\textwidth}
        \centering
        \includegraphics[width=\textwidth]{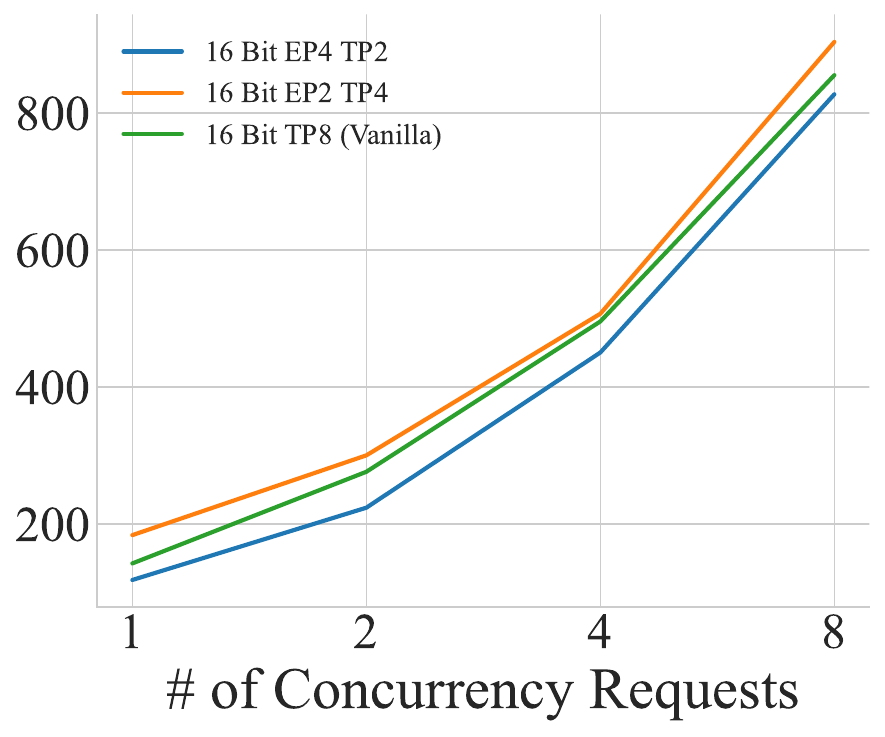}
        \caption{Low conc. with 1 replica.}
        \label{fig:low_concurrency_replica1}
    \end{subfigure}%
    \hfill
    \begin{subfigure}{0.2485\textwidth}
        \centering
        \includegraphics[width=\textwidth]{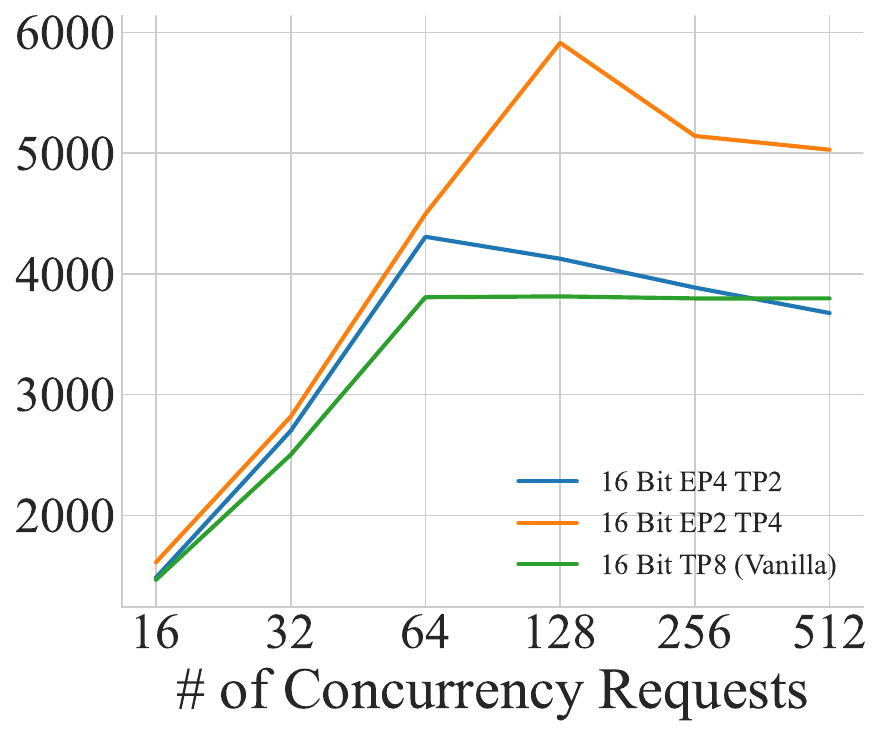}
        \caption{High conc. with 1 replica. }
        \label{fig:high_concurrency_replica1}
    \end{subfigure}%

    \caption {Throughputs for different replica settings and varying \# of concurrency (conc) requests for batch size 64.}
    \label{fig:moe_concurrency_replica}
\end{figure*}

\noindent\textbf{Exp4. Parallelism Comparisons.} We experiment with replica and computations parallelism. For computation parallelism, we test three combinations: Vanilla Tensor Parallelism 8, MOE Expert Parallelism 4 with Tensor Parallelism 2, and MOE Expert Parallelism 2 with Tensor Parallelism 4. We present results in Figure~\ref{fig:moe_concurrency_replica}.

We first evaluate the impact of combining replica parallelism with tensor parallelism to provide through assessment of performance under different parallelism strategies. Specifically, we tested the following configurations using 8-bit quantized Mixtral 8x7B model:
\textit{i}) One replica with Tensor Parallelism 8 (TP8), utilizing 8 GPUs for a single replica \textit{ii}) Two replicas with Tensor Parallelism 4 (TP4), utilizing 4 GPUs per replica; and \textit{iii}) Four replicas with Tensor Parallelism 2 (TP2), utilizing 2 GPUs per replica. These configurations were chosen to equalize utilization of the computational resource for each setup, ensuring a comprehensive but fair evaluation.

As illustrated in Figure ~\ref{fig:low_concurrency_replica124}, at lower concurrency levels, fully utilizing the available compute for tensor parallelism, without any replica parallelism demonstrates superior performance compared to configurations combining tensor and replica parallelism. However, as shown in Figure ~\ref{fig:high_concurrency_replica124}, the trend shifts significantly at higher concurrency levels, favoring configurations with higher degrees of replica parallelism. Notably, the configuration with four replicas and Tensor Parallelism 2 (TP2) significantly outperforms both the two-replica TP4 and single-replica TP8 configurations. Specifically, the four-replica TP2 setup achieves markedly high throughput as the concurrency level exceeds 128 requests while the single-replica TP8 configuration exhibits the poorest performance. The two-replica TP4 configuration shows a modest improvement over the singe-replica TP8 configuration. This study highlights the importance of replica parallelism for handling high concurrency levels, and conversely, highlights the effectiveness of tensor parallelism at lower concurrency levels.

\begin{figure*}[ht]
    \centering
    \includegraphics[width=1\linewidth]{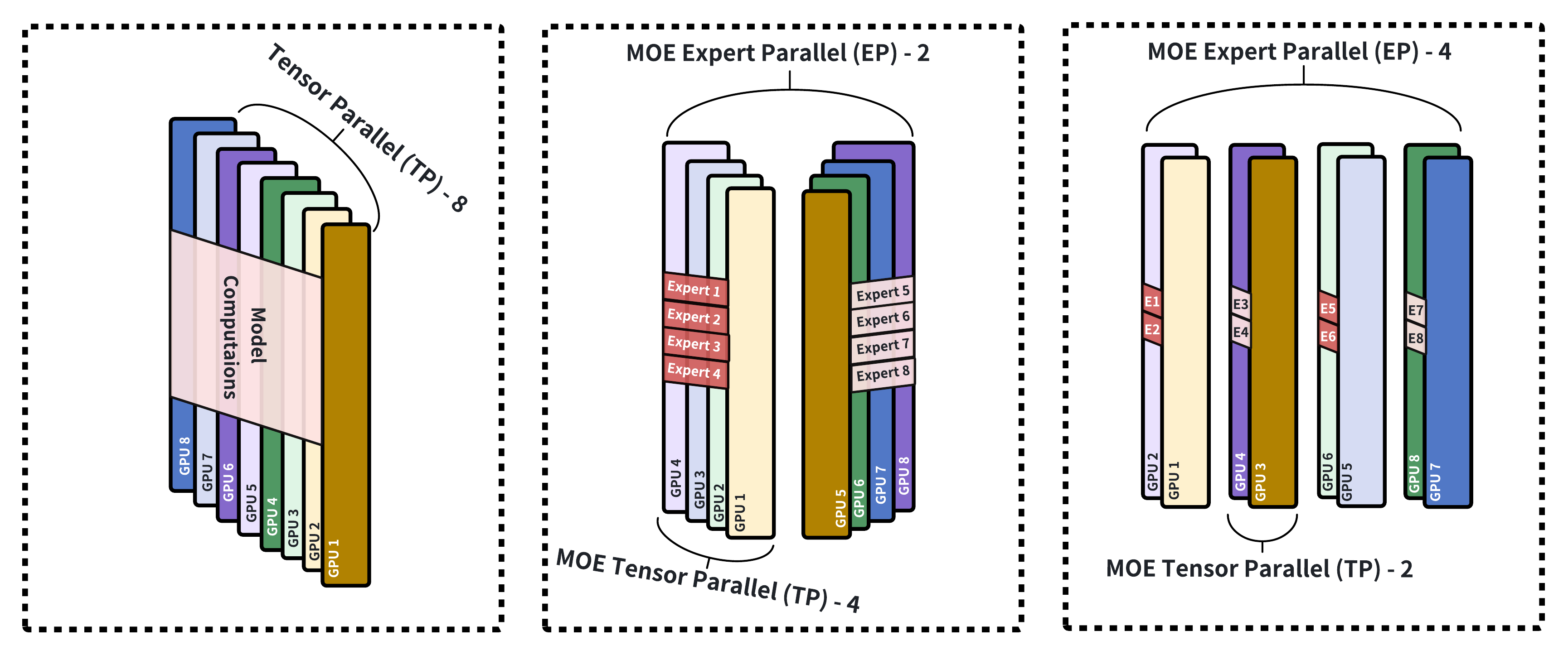}
    \caption{Tensor Parallel and Expert Parallel for Mixture of Experts LLMs.}
    \label{fig:tp_ep_parallelism}
\end{figure*}

We then conducted a series of experiments to assess the performance of variety of computation parallelism techniques as depicted in Figure~\ref{fig:tp_ep_parallelism}. The tested configurations, using a Mixtral 8x7B model include: \textit{i}) Vanilla Tensor Parallelism 8 (TP8) \textit{ii}) MOE Expert Parallelism 4 (EP4) with Tensor Parallelism 2 (TP2); and \textit{iii}) MOE Expert Parallelism 2 (EP2) with Tensor Parallelism 4 (TP4)

Our findings illustrated in Figure ~\ref{fig:low_concurrency_replica1} and ~\ref{fig:high_concurrency_replica1} indicate that MOE-EP2-TP4 consistently outperformed all other methods across the entire concurrency spectrum, demonstrating a particularly significant advantage at higher concurrency levels, specifically beyond 128 concurrent requests. While TP8 showed superior performance compared to MOE-EP4-TP2 at lower concurrency levels, it was eventually surpassed by MOE-EP4-TP2 as concurrency increased beyond 16 requests.

These results underscore the effectiveness of MOE-EP2-TP4 in managing high concurrency scenarios, establishing it as the optimal configuration for deployments intended to handle large-scale concurrency.

\section{Blueprint Architecture of Dynamic Inference Load Balancing System
}\label{sec:blueprint}
\begin{figure}[ht]
    \centering
    \includegraphics[width=1.0\linewidth]
    {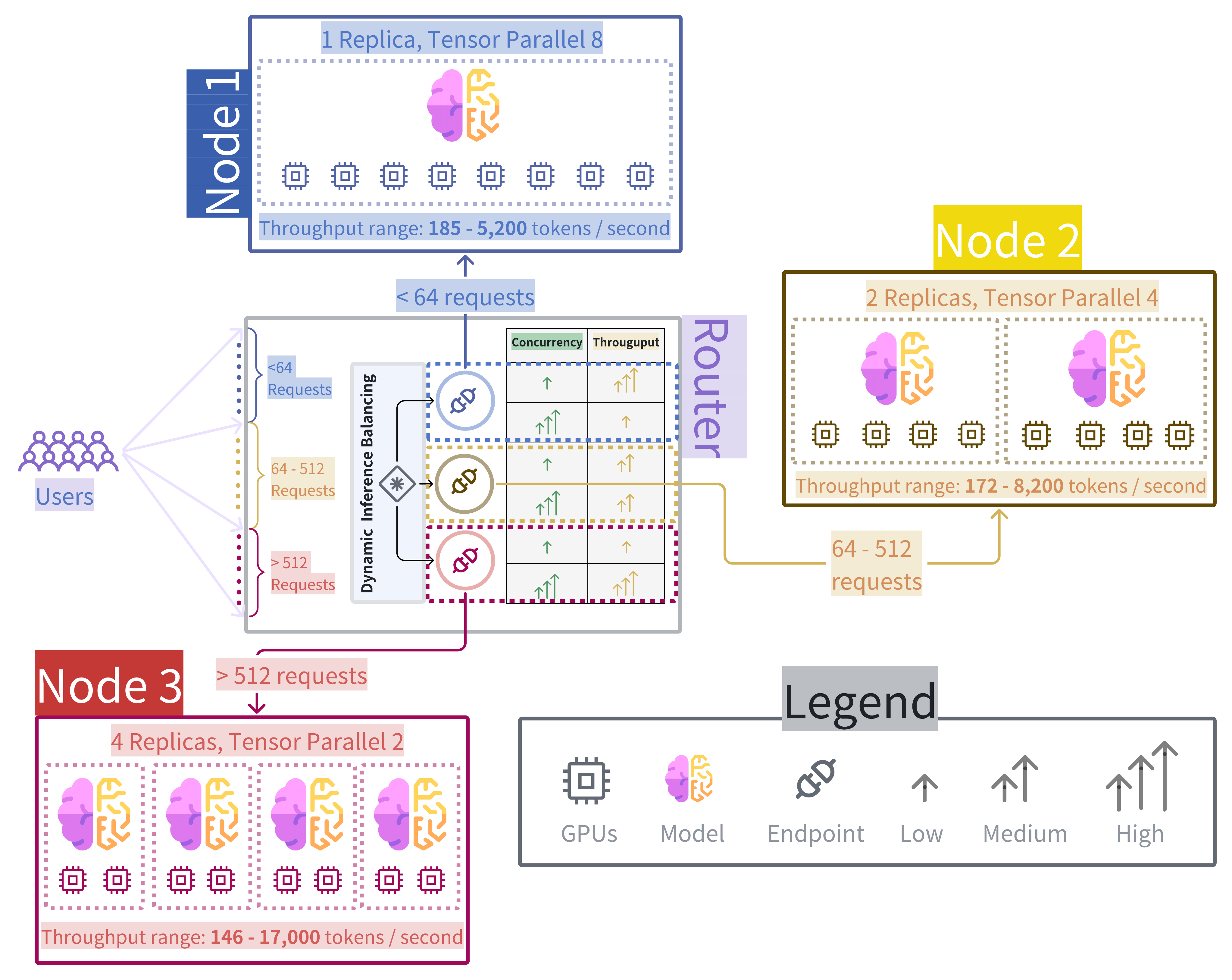}
    \caption{Blueprint Architecture of Dynamic Inference Load Balancing System.}
    \label{fig:dynamic_routing}
    \BBB
\end{figure}
Our experiments have revealed that different engine parameters are suited for different throughput loads, thereby emphasizing the need for a dynamic load balancing system for AI inference unifying the strengths of these heterogeneous configurations and averaging out weakness. We propose a blueprint for such a dynamic inference load balancing system, designed to optimize resource allocation by efficiently distributing inference requests across these heterogeneous replicas, thereby maintaining consistently high throughput regardless of the concurrency scale.

The core component of the proposed system is a dynamic inference balancing router that handles incoming inference requests and intelligently routes them to the appropriate replica based on a routing policy, mapping request concurrency levels to throughput ranges and selecting the  replica best suited to manage the specific workload range.
The dynamic routing policy illustrated in Figure~\ref{fig:dynamic_routing}, showcasing the blueprint architecture and the policy breakdown follows a general rule of thumb:

\noindent\textbf{Low concurrency ($<64$  requests).} Route requests to nodes with fewer replicas but higher tensor parallelism to optimize resource utilization for smaller batch computations.

\noindent\textbf{High concurrency ($\geq64$ requests).} Route requests to nodes with more replicas but lower tensor parallelism, effectively distributing the workload to squeeze everything out of available compute by leveraging the power of replica parallelism.

\section{Conclusion and Future work}

We optimized both engine and platform. With the growing complexity of LLM applications, the platform latency will be more and more important. Instead of optimizing the local inference speed, the industrial research should focus more on simplifying the serving gateway and optimizing the platform.

\bibliographystyle{plainnat}
\bibliography{ref}

\end{document}